# Performance of a nonempirical exchange functional from the density matrix expansion: comparative study with different correlation

Yuxiang Mo[1], Guocai Tian[1,2], and Jianmin Tao[1,*]

[1]Department of Physics, Temple University, Philadelphia, Pennsylvania 19122, USA
[2]State Key Laboratory of Complex Nonferrous Metal Resources Clean Utilization, Kunming University of Science and Technology, Kunming 650093, China

**Abstract**

Recently, Tao and Mo proposed an accurate meta-generalized gradient approximation for the exchange-correlation energy. The exchange part is derived from the density matrix expansion, while the correlation part is obtained by improving the TPSS correlation in the low-density limit. To better understand this exchange functional, in this work, we combine the TM exchange with the original TPSS correlation, which we call TMTPSS, and make a systematic assessment on molecular properties. The test sets include the 223 G3/99 enthalpies of formation, 58 electron affinities, 8 proton affinities, 96 bond lengths, 82 harmonic frequencies, and 10 hydrogen-bonded molecular complexes. Our calculations show that the TMTPSS functional is competitive with or even more accurate than TM functional for some properties. In particular, it is the most accurate nonempirical semilocal DFT for the enthalpies of formation and harmonic vibrational frequencies, suggesting the robustness of TM exchange.

## 1. Introduction

Kohn-Sham Density Functional Theory (DFT)[1] is a standard and practical theory for electronic structure calculations of molecules and solids. This theory simplifies the complicated many-body interaction as a single-particle problem. In this picture, only the exchange-correlation energy component, which accounts for all many-body effects, must be approximated as a functional of the electron spin-densities. Such a simplification endows DFT with great computational efficiency. The active pursuit and rapid progress in the development of exchange-correlation functionals[2–17] lead to remarkable improvement of accuracy and great popularity of DFT in electronic structure calculations.

Consistently accurate exchange-correlation functionals can be constructed with constraint satisfaction approach,[18] or from the underlying hole.[3,19–21] Many universal exact constraints on the exchange-correlation energy have been discovered, including those summarized in Ref. 22. The question is how to properly put them into a density functional. In principle, the more exact constraints a density functional can satisfy, the more consistent accuracy will be expected. To incorporate more exact constraints in a DFT method, more local ingredients have to be used. According to the types of the local ingredients, density functional approximations can be classified into different levels. The simplest one is the local spin-density approximation[23,24] (LSDA), which only makes use of the electron spin-densities to determine the exchange-correlation energy. The next higher level is called the generalized gradient approximation (GGA)[25,26,2,27,4,28–32], which uses not only the electron spin-densities, but also the density gradients as inputs. Density functionals at the meta-GGA level[5,20,33,34] further take into account the Kohn-Sham orbital kinetic energy densities. Each type of local ingredients not only enables better treatment of the one-electron region but also liberates the functional form so that more exact constraints can be satisfied. Inclusion of nonlocal information, such as the exact exchange energy density, leads to nonlocal functionals,[10,35–41] which are more complicated in form and more expensive in computational cost, but can be highly accurate for the description of band gaps,[39,42] charge transfer, reaction barriers, and so forth.

Recently, Tao and Mo (TM) have developed a meta-GGA exchange-correlation functional. The exchange part was obtained from the density matrix expansion. The correlation part is based on the TPSS correlation hole,[43] which is free of self-interaction for any one-electron density and correct for slowly-varying densities, but with a modification in the strong-interaction limit. This modification considerably improves[21] the low-density limit of TPSS functional. Our extensive tests show that this meta-GGA functional can yield very accurate prediction of diverse properties of molecules [44] and solids.[45] To have a better understanding of TM functional with the exchange and correlation parts developed separately, in this work, we combine the TM exchange with the original TPSS correlation, which we call TMTPSS, and assess this combination on standard test sets of molecules and hydrogen-bonded complexes. Our calculations show that, like TM functional, TMTPSS also improves upon DFT methods such as LSDA, PBE GGA, and TPSS meta-GGA for most properties considered here. In particular, it provides the best description of harmonic vibrational frequencies among the semilocal DFTs considered, suggesting the robustness of TM exchange.

## 2. Computational Method

In general, a semilocal functional can be written as form

$$E_x[n] = \int d^3r\, n\, \epsilon_x^{\text{unif}}(n) F_x(n, \nabla n, \tau), \tag{1}$$

where $n$ is the electron density, $\epsilon_x^{\text{unif}}(n)$ is the exchange energy per electron for the uniform electron gas given by $\epsilon_x^{\text{unif}}(n) = -3(3\pi^2 n)^{1/3}/4\pi$, $F_x$ is the enhancement factor, and $\tau(\mathbf{r}) = \sum_{i=1}^{N/2} |\nabla \phi_i(\mathbf{r})|^2$ is the Kohn-Sham kinetic energy density. The specific form of the TMTPSS exchange-correlation energy per electron has been documented in the literature. It is not repeated here. There are different ways to assess density functionals. In the present work, we evaluate the performance of the TMTPSS functional on energetic and structural properties of molecular systems. Our test includes 223 standard enthalpies of formation, 58 electron affinities, 8 proton affinities, T-96R bond lengths, T-82F vibrational frequencies, and H-bond dissociation energies and bond lengths. Calculations are performed with a locally modified Gaussian 09 program.[46] Because our test set includes hydrogen-bonded complexes, to ensure that our assessment is reliable, the large basis set 6-311++G(3*df*,3*pd*) was used in all the calculations except for the enthalpy of formation. (See discussion below for the calculation of enthalpy of formation) The choice of this basis set also allows us to make comparison of our calculations with other DFT data reported in the literature. We use ultrafine grids by setting Grid=UltraFine for the evaluation of two-electron integrals and their derivatives. Geometric optimizations of all molecular systems were carried out with the tight option (Opt=Tight) for cutoffs on forces and step size of geometry variation.

### 3. Results and Discussion

#### 3.1 Standard enthalpies of formation

The standard enthalpy of formation is the enthalpy change in a chemical reaction (at 101.3 kPa) producing one mole of a compound from the most stable elementary substances in standard states. In this work, we adopt the method and experimental atomic data described by Curtiss *et al.*[47] to calculate standard enthalpies of formation at 298 K ($\Delta_f H^{\text{o}}_{298}$). To make direct comparison with other DFT methods reported in the literature, we follow the procedure of Staroverov *et al.*[48] We first calculate the geometries, zero-point energies (ZPE), and thermal corrections using a smaller basis set at the B3LYP/6-31G(2*df*,*p*) level with a frequency scale factor of 0.9854. Then a much larger basis set 6-311++G(3*df*,3*pd*) was adopted to calculate the enthalpy of formation.

As shown in Table I, the TMTPSS functional is the best performing functional among the methods considered for standard enthalpies of formation of G2/148, G3/75, and their combination (G3/99). Interestingly, the MAE of the TMTPSS functional decreases with increasing molecular size from G2 to G3. This is similar to TPSS but in contrast to LSDA, PBE, and TM. This clearly suggests the significance of correlation in the evaluation of enthalpy of formation.

TABLE I. Summary of deviations of the calculated $\Delta_f H^o_{298}$ (kcal/mol) from experiments. The geometries and ZPE were obtained at the B3LYP/6-31G(2df,p) level, with a frequency scale factor of 0.9854. Results of LSDA, PBE, and TPSS functionals are taken from Ref. 48. And those for TM are from Ref. 44. ME = mean error, and MAE = mean absolute error. The smallest and largest MAEs are in bold blue and red, respectively.

| Method | G2 set/148 | | | | G3 set/75 | | | | G3/99 223 | |
|---|---|---|---|---|---|---|---|---|---|---|
| | ME | MAE | Max(+) | Max(-) | ME | MAE | Max(+) | Max(-) | ME | MAE |
| LSDA | -83.7 | **83.7** | 0.4 ($Li_2$) | -207.7 ($C_6H_6$) | -197.1 | **197.1** | None | -347.5 (azulene) | -121.9 | **121.9** |
| PBE | -16.1 | 16.9 | 10.8 ($Si_2H_6$) | -50.5 ($C_3F_4$) | -32.8 | 32.8 | None | -79.7 (azulene) | -21.7 | 22.2 |
| TPSS | -5.2 | 6.0 | 16.2 ($SiF_4$) | -22.9 ($ClF_3$) | -5.2 | 5.5 | 7.5 ($PF_5$) | -12.8 ($S_2Cl_2$) | -5.2 | 5.8 |
| TM | -2.6 | 6.8 | 37.0 ($Si_2H_6$) | -26.6 ($NF_3$) | -2.8 | 9.6 | 22.0 [$Si(CH_3)_4$] | -26.8 ($P_4$) | -2.3 | 7.6 |
| TMTPSS | -4.2 | **5.9** | 37.7 ($Si_2H_6$) | -26.6 ($NF_3$) | -4.6 | **5.2** | 22.0 [$Si(CH_3)_4$] | -26.9 ($P_4$) | -4.4 | **5.6** |

**3.2 Electron affinities**

The electron affinity (EA) characterizes the tendency of a neutral atom or molecule to accept an electron. In a chemical sense, it reflects the acidity of such an atom or molecule in the gaseous environment. The quantity of EA is calculated by subtracting the ZPE-corrected total energy of an atom or molecule at 0 K from that of its corresponding anion. In the evaluation of the TMTPSS functional, the 6-311++G(3df,3pd) basis set was used for the calculation of geometries, electronic energies, and ZPE of both the neutral and anion species. Listed in Table II are the statistic errors of EAs calculated using TMTPSS along with those[44,48] of other functionals.

TABLE II. Summary of deviations from experiment for EAs (eV) of the G3/99 (58 species) test set. We used the 6-311++G(3*df*,3*pd*) basis set to evaluate the geometries, electronic and zero-point energies of TMTPSS. Results of LSDA, PBE, and TPSS functionals are from Ref. 48. Values for the TM functional are from Ref. 44. ME = mean error, and MAE = mean absolute error. The smallest and largest MAEs are in bold blue and red, respectively.

| Method | ME | MAE | Max (+) | Max (-) |
|---|---|---|---|---|
| LSDA | 0.23 | **0.24** | 0.88 ($C_2$) | -0.15 ($NO_2$) |
| PBE | 0.06 | **0.12** | 0.78 ($C_2$) | -0.29 ($NO_2$) |
| TPSS | -0.02 | 0.14 | 0.82 ($C_2$) | -0.32 ($NO_2$) |
| TM | -0.12 | 0.18 | 0.74 ($C_2$) | -0.45 (HOO) |
| TMTPSS | -0.05 | 0.14 | 0.81 ($C_2$) | -0.34 (NCO) |

As seen from Table II, the TMTPSS functional also underestimates EAs, similar to TPSS and TM meta-GGAs. In terms of MAE, the TMTPSS functional is in par with TPSS, both with an MAE of 0.14 eV, smaller than those of the TM (MAE=0.18 eV) and LSDA (MAE=0.24 eV), but larger than that of PBE (MAE=0.12 eV). However, the accuracy of the TMTPSS and other exchange-correlation functionals cannot be judged based solely on the errors they yield, as such errors may include contributions from cancellation of the electron self-repulsion and suppression of the unbound states with finite basis set.[49] The largest overestimation of EA by TMTPSS is also found for the $C_2$ molecule. This is a result of the inadequate description of the singlet ground state of the $C_2$ molecule by a single determinant.[50–52]

**3.3 Proton affinities**

Proton affinity (PA) is the amount of energy released in the process of adding a proton to a species at its ground-state. This quantity describes the ability of a species to accept a proton, thereby hinting the gas-phase basicity of such a species. Deviations of calculated PAs of the 8 species of the G3/99 test set from experiments are summarized in Table III. ZPEs are included during the evaluation of the energies of both the pristine species and its protonated counterpart. The TMTPSS overestimates PAs for the 8 species, with an ME of 1.5 eV. Its MAE of 1.8 eV is the same as that of the TPSS, which is larger than the MAEs of TM (1.2 eV) and PBE (1.6 eV), but smaller than that of the LSDA (MAE=5.9 eV).

TABLE III. Summary of deviations from experiments of PAs (eV) for the G3/99 (8 species) test set. Results of LSDA, PBE, and TPSS functionals are taken from Ref. 48, while those of TM functional are from Ref. 44. We used the 6-311++G(3$df$,3$pd$) basis set to evaluate the geometries, electronic and zero-point energies of TMTPSS. ME = mean error, and MAE = mean absolute error. The smallest and largest MAEs are in bold blue and red, respectively.

| Method | ME | MAE | Max (+) | Max (-) |
|--------|------|------|---------------|---------------|
| LSDA   | -5.9 | **5.9** | None | -10.6 ($PH_3$) |
| PBE    | -0.8 | 1.6  | 2.4 ($C_2H_2$) | -3.6 ($PH_3$) |
| TPSS   | 1.7  | 1.8  | 4.4 ($C_2H_2$) | -0.5 ($H_2O$) |
| TM     | 0.7  | **1.2** | 4.3 ($C_2H_2$) | -1.5 ($H_2O$) |
| TMTPSS | 1.5  | 1.8  | 4.9 ($C_2H_2$) | -1.3 ($H_2O$) |

### 3.4 Bond lengths

We adopted the T-96R test set[48] of 96 ground-state molecules (10 molecular cations and 86 neutral molecules) to assess the accuracy of TMTPSS in predicting equilibrium molecular bond length, which is a deciding quantity for electronic structure and other properties. Listed in Table IV is the summary of deviations for calculated equilibrium bond lengths using TMTPSS, in comparison with other functionals. The experimental values of equilibrium bond lengths are from Ref. 53 for $Be_2$, Ref. 54 for NaLi and cations, and Ref. 55 for the rest. As shown in Table IV, the TMTPSS has an MAE of 0.013 Å for T-96R bond lengths, which is smaller than those of PBE (MAE=0.016 Å) and TPSS (MAE=0.014 Å), equal to that of LSDA, and larger than that of TM (MAE=0.012 Å).

TABLE IV. Summary of deviations from experiments of equilibrium bond lengths (Å) for the T-96R test set. Results of LSDA, PBE, and TPSS functionals are taken from Ref. 48. Values of LSDA do not include $F_2^+$ and SF (fails to converge). Results of the TM functional are from Ref. 44. The TMTPSS results were calculated using the basis set 6-311++G(3*df*,3*pd*). ME = mean error, and MAE = mean absolute error. The smallest and largest MAEs are in bold blue and red, respectively.

| Method | ME | MAE | Max (+) | | Max (-) | |
|---|---|---|---|---|---|---|
| LSDA | 0.001 | 0.013 | 0.042 | (BN) | -0.094 | ($Na_2$) |
| PBE | 0.015 | **0.016** | 0.055 | ($Li_2$) | -0.013 | ($Be_2$) |
| TPSS | 0.014 | 0.014 | 0.078 | ($Li_2$) | -0.008 | ($P_4$) |
| TM | 0.010 | **0.012** | 0.054 | ($Li_2$) | -0.086 | ($Si_2$) |
| TMTPSS | 0.013 | 0.013 | 0.060 | ($Li_2$) | -0.009 | ($P_4$) |

### 3.5 Harmonic vibrational frequencies

The harmonic vibrational frequency can be used to identify molecular structure via infrared spectroscopy. To evaluate the accuracy of TMTPSS on harmonic vibrational frequencies, we used the T-82F test set[48] of 82 ground-state diatomic molecules (69 neutral species and 13 cations). Listed in Table V are the deviations of calculated harmonic vibrational frequencies from experimental data. The experimental values are from Ref. 53 for $Be_2$, Ref. 54 for NaLi and cations, and Ref. 55 for the others. According to Table V, the TMTPSS functional also underestimates the harmonic vibrational frequencies, with an ME of -16.6 cm$^{-1}$. The TMTPSS functional has an MAE of 29.0 cm$^{-1}$, which is slightly smaller than those of the meta-GGAs TM (MAE=29.7 cm$^{-1}$) and TPSS (MAE=30.4 cm$^{-1}$), but significantly improves upon GGA PBE (MAE=42.0 cm$^{-1}$), and LSDA (MAE=48.9 cm$^{-1}$).

TABLE V. Summary of deviations from experiments of harmonic vibrational frequencies ($\omega_e$) in cm$^{-1}$ of the T-82F (82 diatomic molecules) test set. Results of LSDA, PBE, and TPSS functionals are taken from Ref. 48. Values of LSDA do not include $F_2^+$ (fails to converge). Results of the TM functional are from Ref. 44. In the evaluation of the TMTPSS functional, we used the 6-311++G(3$df$,3$pd$) basis set for the calculation of both the geometries and harmonic vibrational frequencies. ME = mean error, and MAE = mean absolute error. The smallest and largest MAEs are in bold blue and red, respectively.

| Method | ME | MAE | Max (+) | Max (-) |
|--------|------|------|---------------|---------------|
| LSDA   | -11.8 | **48.9** | 140.7 ($F_2$) | -227.7 ($H_2$) |
| PBE    | -31.7 | 42.0 | 82.5 ($Be_2$) | -175.3 ($HF^+$) |
| TPSS   | -18.7 | 30.4 | 81.2 ($F_2^+$) | -145.9 (HF) |
| TM     | -13.5 | 29.7 | 91.4 ($F_2^+$) | -145.2 (HF) |
| TMTPSS | -16.6 | **29.0** | 80.3 ($F_2^+$) | -157.1 (HF) |

### 3.6 Hydrogen-bonded complexes

Accurate description of weakly-bonded systems is particularly important for biochemical processes. It determines the second-order configuration of biomolecular chains (e.g., DNA double helix structure) that define biological activities. In this work, we use the test set of Rabuck and Scuseria[56] to evaluate the TMTPSS functional. The tested systems include 10 pairs: $(HF)_2$, $(HCl)_2$, $(H_2O)_2$, HF/HCN, HF/H$_2$O, CN$^-$/H$_2$O, OH$^-$/H$_2$O, HCC$^-$/H$_2$O, H$_3$O$^+$/H$_2$O, and NH$_4^+$/H$_2$O.

Illustrations of bond lengths for these pairs are available in Fig. 1 of Ref. 56. Summarized in Table VI are the deviations of calculated 10 dissociation energies ($D_0$) and 11 H-bond lengths. From Table VI, we can also observe that the TMTPSS functional has an MAE of 0.6 kcal/mol which is equal to that of TPSS. It is smaller than those of LSDA (MAE=5.8 kcal/mol) and PBE (MAE=1.0 kcal/mol), but larger than that of TM (MAE=0.3 kcal/mol). In terms of H-bond dissociation energies, the TMTPSS functional has an MAE of 0.041 Å, being less accurate than TM (MAE=0.017 Å) and TPSS (MAE=0.021 Å) but more accurate than LSDA (MAE= 0.147 Å) and PBE (MAE=0.043 Å).

TABLE VI. Statistic errors of bond lengths (Å) and ZPE-corrected dissociation energies $D_0$ (kcal/mol) of 10 hydrogen-bonded complexes. The 6-311++G(3*df*,3*pd*) basis set is adopted for the TMTPSS calculation of the geometries, electronic and zero-point energies. Results of LSDA, PBE, and TPSS functionals are taken from Ref. 48. And those for TM are from Ref. 44. We use the MP2(full)/6-311++G(3*df*,3*pd*) results[48] as the reference in the evaluation of errors. ME = mean error, and MAE = mean absolute error. The smallest and largest MAEs are in bold blue and red, respectively.

|        | $D_0$ (kcal/mol) |       | Bond lengths (Å) |       |
|--------|------|------|--------|-------|
| Method | ME   | MAE  | ME     | MAE   |
| LSDA   | 5.8  | **5.8** | -0.127 | **0.147** |
| PBE    | 0.9  | 1.0  | -0.018 | 0.043 |
| TPSS   | 0.3  | 0.6  | -0.006 | 0.021 |
| TM     | -0.1 | **0.3** | 0.014  | **0.017** |
| TMTPSS | -0.6 | 0.6  | 0.037  | 0.041 |

## 4. Conclusions

In conclusion, we have made a comparative assessment of TMTPSS functional on molecular systems. The TMTPSS functional has the best accuracies of all functionals considered on G3/99 standard enthalpies of formation and harmonic vibrational frequencies. For electron affinities, the TMTPSS functional is more accurate than TM and in par with meta-GGA TPSS. For proton affinities and bond lengths, the TMTPSS functional is less accurate than TM but comparable with TPSS. The TMTPSS functional has larger errors than TM for hydrogen-bonded complexes, but in terms of dissociation energy it is as accurate as TPSS. Overall, our results show that the TMTPSS meta-GGA functional is one of the most accurate functionals and therefore a great choice for studying properties of molecules, especially standard enthalpies of formation and harmonic vibrational frequencies.

**Acknowledgments**

JT and YM acknowledge support from the NSF under Grant No. CHE 1640584. JT also acknowledges support from Temple University. GT was supported by China Scholarship Council and the National Natural Science Foundation of China under Grant No. 51264021. Computational support was provided by the HPC of Temple University.

* Corresponding author. jianmin.tao@temple.edu

**References：**

[1] W. Kohn and L. J. Sham, Phys. Rev. **140**, A1133 (1965).
[2] C. Lee, W. Yang, and R. G. Parr, Phys. Rev. B **37**, 785 (1988).
[3] A. D. Becke and M. R. Roussel, Phys. Rev. A **39**, 3761 (1989).
[4] J. P. Perdew, K. Burke, and M. Ernzerhof, Phys. Rev. Lett. **77**, 3865 (1996).
[5] J. Tao, J. P. Perdew, V. N. Staroverov, and G. E. Scuseria, Phys. Rev. Lett. **91**, 146401 (2003).
[6] J. Heyd, G. E. Scuseria, and M. Ernzerhof, J. Chem. Phys. **118**, 8207 (2003).
[7] R. Armiento and A. E. Mattsson, Phys. Rev. B **72**, 085108 (2005).
[8] Y. Zhao and D. G. Truhlar, J. Phys. Chem. A **110**, 13126 (2006).
[9] J. P. Perdew, A. Ruzsinszky, G. I. Csonka, O. A. Vydrov, G. E. Scuseria, L. A. Constantin, X. Zhou, and K. Burke, Phys. Rev. Lett. **100**, 136406 (2008).
[10] Y. Zhao and D. G. Truhlar, Theor. Chem. Acc. **120**, 215 (2008).
[11] J. P. Perdew, A. Ruzsinszky, G. I. Csonka, L. A. Constantin, and J. Sun, Phys. Rev. Lett. **103**, 026403 (2009).
[12] R. Peverati and D. G. Truhlar, J. Phys. Chem. Lett. **2**, 2810 (2011).
[13] R. Peverati and D. G. Truhlar, Phys. Chem. Chem. Phys. **14**, 16187 (2012).
[14] A. V. Arbuznikov and M. Kaupp, J. Chem. Phys. **141**, 204101 (2014).
[15] J. Sun, A. Ruzsinszky, and J. P. Perdew, Phys. Rev. Lett. **115**, 036402 (2015).
[16] H. S. Yu, W. Zhang, P. Verma, X. He, and D. G. Truhlar, Phys. Chem. Chem. Phys. **17**, 12146 (2015).
[17] H. S. Yu, X. He, S. L. Li, and D. G. Truhlar, Chem. Sci. **7**, 5032 (2016).
[18] J. P. Perdew, A. Ruzsinszky, J. Tao, V. N. Staroverov, G. E. Scuseria, and G. I. Csonka, J. Chem. Phys. **123**, 062201 (2005).
[19] J. P. Perdew and W. Yue, Phys. Rev. B **33**, 8800 (1986).
[20] T. Van Voorhis and G. E. Scuseria, J. Chem. Phys. **109**, 400 (1998).
[21] J. Tao and Y. Mo, Phys. Rev. Lett. **117**, 073001 (2016).
[22] V. N. Staroverov, G. E. Scuseria, J. Tao, and J. P. Perdew, Phys. Rev. B **69**, 075102 (2004).
[23] S. H. Vosko, L. Wilk, and M. Nusair, Can. J. Phys. **58**, 1200 (1980).
[24] J. P. Perdew and Y. Wang, Phys. Rev. B **45**, 13244 (1992).
[25] J. P. Perdew, Phys. Rev. B **33**, 8822 (1986).
[26] A. D. Becke, Phys. Rev. A **38**, 3098 (1988).
[27] J. P. Perdew, J. A. Chevary, S. H. Vosko, K. A. Jackson, M. R. Pederson, D. J. Singh, and C. Fiolhais, Phys. Rev. B **46**, 6671 (1992).
[28] R. H. Hertwig and W. Koch, Chem. Phys. Lett. **268**, 345 (1997).
[29] F. A. Hamprecht, A. J. Cohen, D. J. Tozer, and N. C. Handy, J. Chem. Phys. **109**, 6264 (1998).
[30] M. Ernzerhof and G. E. Scuseria, J. Chem. Phys. **110**, 5029 (1999).
[31] A. D. Boese and N. C. Handy, J. Chem. Phys. **114**, 5497 (2001).
[32] N. C. Handy and A. J. Cohen, Mol. Phys. **99**, 403 (2001).
[33] J. P. Perdew, S. Kurth, A. Zupan, and P. Blaha, Phys. Rev. Lett. **82**, 2544 (1999).
[34] Y. Zhao and D. G. Truhlar, J. Chem. Phys. **125**, 194101 (2006).
[35] J. Jaramillo, G. E. Scuseria, and M. Ernzerhof, J. Chem. Phys. **118**, 1068 (2003).
[36] J. P. Perdew, V. N. Staroverov, J. Tao, and G. E. Scuseria, Phys. Rev. A **78**, 052513 (2008).
[37] R. Haunschild, B. G. Janesko, and G. E. Scuseria, J. Chem. Phys. **131**, 154112 (2009).

[38] R. Haunschild and G. E. Scuseria, J. Chem. Phys. **132**, 224106 (2010).

[39] M. A. L. Marques, J. Vidal, M. J. T. Oliveira, L. Reining, and S. Botti, Phys. Rev. B **83**, 035119 (2011).

[40] R. Haunschild, M. M. Odashima, G. E. Scuseria, J. P. Perdew, and K. Capelle, J. Chem. Phys. **136**, 184102 (2012).

[41] A. J. Garza, I. W. Bulik, T. M. Henderson, and G. E. Scuseria, Phys. Chem. Chem. Phys. **17**, 22412 (2015).

[42] F. Tran and P. Blaha, Phys. Rev. Lett. **102**, 226401 (2009).

[43] L. A. Constantin, J. P. Perdew, and J. Tao, Phys. Rev. B **73**, 205104 (2006).

[44] Y. Mo, G. Tian, R. Car, V. N. Staroverov, G. E. Scuseria, and J. Tao, submitted. http://arxiv.org/abs/1607.05249

[45] Y. Mo, R. Car, V. N. Staroverov, G. E. Scuseria, and J. Tao, submitted. http://arxiv.org/abs/1607.05252

[46] M. J. Frisch *et al.*, Gaussian 09, Revision A.02 (Gaussian, Inc., Wallingford CT, 2009).

[47] L. A. Curtiss, K. Raghavachari, P. C. Redfern, and J. A. Pople, J. Chem. Phys. **106**, 1063 (1997).

[48] V. N. Staroverov, G. E. Scuseria, J. Tao, and J. P. Perdew, J. Chem. Phys. **119**, 12129 (2003).

[49] N. Rösch and S. B. Trickey, J. Chem. Phys. **106**, 8940 (1997).

[50] A. Goursot, J. P. Malrieu, and D. R. Salahub, Theor. Chim. Acta **91**, 225 (1995).

[51] P. R. T. Schipper, O. V. Gritsenko, and E. J. Baerends, Theor. Chem. Acc. **99**, 329 (1998).

[52] A. V. Kopylow and D. Kolb, Chem. Phys. Lett. **295**, 439 (1998).

[53] J. M. L. Martin, Chem. Phys. Lett. **303**, 399 (1999).

[54] K. P. Huber and G. Herzberg, *Molecular Spectra and Molecular Structure. IV. Constants of Diatomic Molecules* (Van Nostrand, New York, 1979).

[55] D. R. Lide, *CRC Handbook of Chemistry and Physics*, 83rd ed. (CRC, Boca Raton, FL, 2002).

[56] A. D. Rabuck and G. E. Scuseria, Theor. Chem. Acc. **104**, 439 (2000).